# Comparison of different methods for evaluating the transmission function of a two-stage cylindrical mirror analyzer in XPS applications

Pierre BRACCONI* and Olivier HEINTZ

Institut Carnot de Bourgogne, UMR 5209 CNRS,

Université de Bourgogne, 9 ave. Alain Savary, BP 47870, F-21078 Dijon, France

* corresponding author:

Tel. : + 33 (0)380 396 154. / Fax : + 33 (0)380 396 132

e-mail : pierre.bracconi@u-bourgogne.fr

**Abstract**

Three different evaluations of the energy dependence of the transmission-detection function of a two-stage cylindrical electron analyzer have been obtained by resorting to three established procedures. Their relative merits have been tested as follows. First they have been used to correct raw XPS spectra of clean Cu, Ag and Au surfaces. Next, the secondary electron background has been subtracted using Tougaard's method. Finally, the primary electron spectra so obtained have been reanalyzed by peak area measurement in the frame of the modern formalism for quantitative XPS analysis. Ideally a constant residual value should thus be obtained. The variability of these residuals with peak energy allows an objective rating of the initial transmission-detection functions.

# I - Introduction

Strictly, the application of XPS to the quantitative chemical analysis of surfaces requires the prior knowledge of the transmission-detection function of the electron spectrometer used and the correction thereof of the raw photoelectron spectra, I(E), to yield the true spectrum, n(E), emitted from the analyzed surface.
In this context the first principles method (FPM) based on the modern formalism for XPS analysis and photoelectron peak area measurement would be ideally suited to the determination of the overall the transmission-detection function of the spectrometer provided it is applied to a spectrum corrected beforehand for the background of secondary electrons. In the case of metals and alloys this can be achieved with good accuracy by resorting to Tougaard's method [1]. Unfortunately, that method can rigorously be applied only to the *true* spectrum itself. This is why in practice the FPM has been and may still be inappropriately applied to raw photoelectron spectra of metals simply corrected for the background of secondary photoelectrons using Shirley's method [2]. And this is also why alternative methods for the direct the determination of the transmission-detection function of electron spectrometers have been developed. They include:

- Modeling and/or simulation of transmission and scattering of electrons through the collector grids and analyzer and of the efficiency of the electron detector. In the case of cylindrical mirror analyzers (CMA) the publications of Staib [3] and Staib and Dinklage [4] failed to yield an explicit equation applicable to practical experimental conditions with the ISA Riber MAC2 analyzer used in the present work.
- Comparison method. In short this consists in dividing a raw experimental spectrum by the true spectrum of the same element i.e. a spectrum corrected for the known transmission function of the spectrometer used for its acquisition [5]. Here we will use the National Physical Laboratory (NPL) photoelectron spectrometer intensity calibration system [6].
- The so called bias method [7] which to the authors' knowledge have never been utilized to determine of the transmission-detection function of any CMA-based electron spectrometer.

As for the particular case of the ISA Riber MAC2 analyser (shortened to MAC2 in the following) only one publication has been dedicated to the problem: Repoux et alii [8] relied on the basic version of the FPM (background subtraction from raw spectra by Shirley's method prior to peak area measurement and basic formalism for XPS analysis) to obtain discrete values of the overall transmission-detection function, $TD(E)$, from Cu and Ag spectra with two X-ray sources (Al and Mg anticathodes). They proposed the following power function to represent its dependence on photoelectron kinetic energy $E_K$ and analyzer pass energy value $E_{pass}$ (focusing voltage VF1 applied to the retarding grid) :

[1] $TD(E_k) = KE_k^n$ with $K = 211 - 10.8 VF_1$ and $n = 1.23 - 0.043 VF_1$

However, close examination of the many graphical results reported in this paper shows that the *very same* dispersion (distribution) of the data points around the fit of Eq.[1] is present in *all* figures. In particular, it is almost exactly duplicated (except for the shift of the kinetic energy scale) when the anticathode is changed from Al to Mg, all other parameters being kept unchanged. This is evidence that the observed dispersion is not the result of random measuring errors but rather of *systematic* errors that can be (at least partly) linked with the combined effect of the background subtraction method and internal scattering in the CMA. In support with this we may note that the authors used a monotonously varying power function of kinetic energy to account for the inelastic mean free path values (IMFP), which cannot be the cause of the data points dispersion just mentioned.

Since then much progress has been done in the theory and the modern formalism for quantitative XPS analysis that accounts for elastic electron scattering has evolved. Based on this, it seemed interesting to look for possible improvements in the determination of the transmission function by the first principles method. This modern XPS formalism has been extensively covered in the literature over the last two decades, in particular by A. Jablonski, M. P. Seah and their collaborators . As we make use of it in the final computational step of our method that is to be described next, a summary of the theoretical material and particular considerations regarding its utilization in connection with the MAC2 analyzer are given in Appendix-A and Appendix-B respectively.

The idea that underlies the method developed in the present paper may be stated as follows:
1. Cu, Ag and Au XPS spectra corrected for the (NPL Metrology Lab) spectrometer transmission function can be obtained from the NPL software and data package as explained in the experimental section §II-1. They are referred to as the *NPL reference* *spectra* in the following. After subtraction of the secondary electron background using Tougaard's method, they should represent the best possible estimate of the true primary electron spectra (PES) of Cu, Ag and Au.

2. Applying the first principle method to the PES so obtained should ideally generate a set of constant values (one per emission line) *independent* of electron kinetic $E_K$. Each would be a particular evaluation of what we will refer to as the residual transmission function (RTF). As obvious from the theory in Appendix A, it is the product of the true transmission-detection function of the spectrometer with several other technical parameters including the flux of incident X photons.

3. Consider now raw experimental spectra of clean Cu, Ag and Au reference samples, acquired with any other spectrometer of interest (MAC2 in the present work). Starting with the NPL standard procedure one generates a first estimate of the transmission-detection function for that particular spectrometer (apart from a constant multiplying factor). Then, by repeating steps 1) and 2) above, one should, ideally, arrive at a constant RTF (essentially the same that would be obtained by similarly processing the NPL reference spectra). In contrast, starting with transmission-detection function evaluated by alternative methods, the same procedure should produce different RTF and thus allow the comparison of their relative merit by simple statistical analysis. Notice that this will also provide a direct evaluation of a lower bound to the accuracy that can be attained when measuring the stoichiometry of a surface compound or the concentrations of the components of a mixture of surface phases: the relative error to be expected could not be lower than the variation coefficient of the RTF obtained in similar experimental conditions.

In the next section, the comparison of the RTF according to step 3 above will be extended to two additional situations. One is the absence the initial correction by a transmission function (equivalent to correction by a unit function), in which case the RTF values should logically exhibit the largest variability. The other will consist in computing the RTF directly from the NPL PES. In that case a nearly perfect constancy is expected and, be any variability observed, it would eventually represent the combined contributions of errors in the NPL PES themselves, Tougaard's background correction and peak area evaluation and also of the fact that the actual instrumental parameters in the NPL PES acquisition are not taken into account in the computations.

## II – Experimental

II-1 Materials and techniques

The reference copper, silver and gold samples were those provided by the National Physical Laboratory for use with their XPS intensity calibration system [**9**]. The analytical procedure and the final carbon and oxygen surface contamination levels complied with the recommendations of the spectrum acquisition guide.

The base pressure achieved in the analytical chamber during spectra acquisition never exceeded $2.10^{-9}$ mbar. The Ar ion gun was operated at 4KeV during 60 min prior spectrum acquisition. The X-ray tube was operated at medium power 240 W with an Al anticathode.

The MAC2 is a two-stage analyzer a descriptive sketch of which can be found in [**8**]. The following details are important in the present context. The input optics (first stage) focuses the analyzed area of the sample surface on a diaphragm that acts as a source for the CMA (second stage). The acceptance solid angle is limited by two coaxial cones whose common axis coincides with the spectrometer revolution axis that is set perpendicular to the sample surface. In the following, we assume their half opening angles $\alpha_{min}$ and $\alpha_{max}$ equal to 24° and 38° respectively, in agreement with the description in [**4**]. Notice that this is at variance with the traditional design of Plamberg [**10**] in which $\alpha = 42.3°(\pm 6°)$ and one consequence is that the transmitted electrons are focused on an off-axis annular slit. The acceptance solid angle in Staib and Dinklage design equals 0.79 sr, i.e. 12.55% of $2\pi$. Though, for the MAC2 the figure is 6% of $2\pi$ (and this is the only technical data available from the manufacturer) and most likely corresponds to a narrower $\alpha$ range: $\Delta\alpha = \pm 3,3°$ instead of $\pm 7°$. That discrepancy has been ignored in the following for it would have no influence on the conclusions. As shown in Appendix-B, the average value of the term $\overline{\cos^2 \psi}$ to be used with Eq.[A7] equals 0.3019 (using Staib's $\alpha$ angle values) and the variation of that quantity with $\Delta\alpha$ are excessively small.

The energy scale calibration was carried out using ion etched Cu and Ag samples with Al-K$\alpha$ and Mg-K$\alpha$ X-ray lines. The peak positions of the Cu-$2p^{3/2}$ and Auger Cu-$L_3$VV lines, Ag$3d^{5/2}$ and Auger Ag-$M_5$VV and Ag-$M_4$VV lines were adjusted to the reference values recommended in [**11**]. The subsequent control showed that the difference between measured and reference values was reduced to -0.06eV at $E_B \approx$ 0eV and +0.04eV at $E_B \approx$ 1000eV.

The MAC2 is operated in the constant ΔE mode and the spectra were acquired with 1eV energy steps, over the range 200 eV < KE < 1560 eV with the channeltron (CEM) working tension set at 2kV.

II-2 Methodology

In a first step one defines several analytical expressions (approximations) of the transmission-detection function with the help of a) the NPL XPS intensity calibration system, b) the empirical equation of Repoux et alii, and c) the "bias method". These points are developed below.

The second step consists in correcting the experimental Cu, Ag and Au spectra for the contribution of these transmission-detection functions (i.e. dividing the experimental signal intensity at a given kinetic energy by the value of the transmission function at the same energy) and proceeding to the background subtraction using Tougaard's method. One thus obtains as many tentative evaluations the PES of of Cu, Ag and Au as different transmission-detection function representations used.

Finally one measures peak areas and computes the RTF values from :

$$[2] \quad RTF = \frac{peak\ area\ of\ PES}{N_V \lambda \cos\alpha \left(\frac{d\sigma}{d\Omega}\right)}$$

One set of RTF values is associated with each transmission-detection function used and is characterized by basic statistical parameters.

Background corrections and peak area measurements were carried out with the help of CasaXPS® software. The background was computed on selected separate sections of the spectra using the C4T equation for the universal (differential inelastic electron) cross-section:

$$[3] \quad K(T) \cong \frac{1}{\lambda} \frac{BT}{(C+T^2)^2}$$

B and C were given the initial values recommended by Tougaard [12], 2866 eV$^2$ and 1643 eV$^2$ respectively. Then, they were varied to arrive at an optimum fit. In all cases the condition B < 2C was imposed [13]. Because more or less variable but equally acceptable results may be obtained depending on the final values of B, C and T, we usually defined upper and lower bounds of peak areas instead of single values. This of course resulted in upper and lower bounds of the RTF.

The $K\alpha_{3,4}$ satellites, about 10eV above the $K\alpha_{1,2}$ peaks, are not subtracted from the raw spectra and are included in the integrated regions, which means that all measured peak area values are increased by about 11% compared to areas of $K\alpha_{1,2}$ lines. In any case, the total areas of p and d lines are obtained by integration. The areas of the 2 components (e.g. $p^{1/2}$ and $p^{3/2}$) are computed from the theoretical proportions (e.g. 1/3, 2/3). However this is no longer possible when the tail of two lines such as Cu-3s and Cu-3p overlap inherently [**14**]. In such situations, only the total area can be obtained by integration but then the RTF value computed from Eq.[4] represents a certain weighted mean value of the two components:

$$[4] \quad RTF = \frac{I^{Cu-3s} + I^{Cu-3p}}{N_V^{Cu} \lambda^{Cu-3s} \cos\alpha \left(\frac{d\sigma^{Cu-3s}}{d\Omega}\right) + N_V^{Cu} \lambda^{Cu-3p} \cos\alpha \left(\frac{d\sigma^{Cu-3p}}{d\Omega}\right)}$$

The result so obtained is exact only if the two components $A = \dfrac{I^{Cu-3s}}{N_V^{Cu} \lambda^{Cu-3s} \cos\alpha \left(\dfrac{d\sigma^{Cu-3s}}{d\Omega}\right)}$, and

$B = \dfrac{I^{Cu-3p}}{N_V^{Cu} \lambda^{Cu-3p} \cos\alpha \left(\dfrac{d\sigma^{Cu-3p}}{d\Omega}\right)}$ behave ideally i.e. are equal. If they are not, the error done (in identifying the RHS of Eq.[4] with the mean (A+B)/2) would range from +200% to -200% when the ratio B/A increases from 0 to infinity, but a ±10% difference of A and B with their mean (A+B)/2 would result in 10% error in the RTF value computed from Eq.[4].

II-2-1 NPL Intensity Calibration System.

In short the method consists in dividing a raw experimental spectrum by a reference (true) spectrum of the same element, i.e. corrected for the "known" transmission function of the spectrometer used for its acquisition, here the NPL metrology spectrometer (MSPEC) [**15, 16**]. Running the calibration software yields an analytic expression of the reduced variable $\varepsilon = E_K/1000$ and statistical criteria to appreciate the reliability of the result. The (default) analytic expression is a polynomial fraction of the form

$$[5] \quad TD_{NPL}(E_K) = \sum_{i=0}^{4} a_i \varepsilon^i \bigg/ \sum_{i=1}^{4} (1 + b_i \varepsilon^i)$$

Processing our Cu, Ag and Au spectra we obtained the following values of the parameters :
$a_0$ = 40.420648, $a_1$ = -145.853870, $a_2$ = 247.325350, $a_3$ = 396.837110, $a_4$ = -93.351564, $b_1$ = -3.136668, $b_2$ = +3.182537, $b_3$ = 18.065769 and $b_4$ = 11.0406791.

Besides, experimental Cu, Ag and Au spectra recorded with the MSPEC by S. Spencer are provided with the NPL calibration system. When processed with the software as any other experimental spectra, they are expected to yield the IERF of that spectrometer, (i.e. $TD_{MSPEC}(E_K)$. The result so obtained is put in the standard analytic form, with the following values for the parameters: $a_0$ = 4361.494000, $a_1$ = -4.066375, $a_2$ = 4.136176, $a_3$ = -1.029355, $a_4$ = -12.748294, $b_1$ = 0.592284, $b_2$ = -0,104131, $b_3$ = -0,014479 and $b_4$ = 0.046936. Finally, using this, the true spectra of the 3 elements Cu, Ag and Au can be computed.

After adjustment of the energy scales (to make peak positions coincide) and subtraction of an offset intensity (mean intensity above $E_K$=1550 eV) the mere division of our experimental spectra by these true spectra should produce the same result as the direct use of NPL software. The interest in doing so is that unsmoothed numerical data are obtained which may help understanding the shortcomings of the present application of the NPL software to the MAC2 analyzer in the present conditions of operation (constant $\Delta E$ mode and low pass energy).

All these data are compared in **Fig.1**. Obviously a non-monotonous decrease is systematically observed on increasing $E_K$. At low $E_K$ the agreement between the *IERF* and the individual divided spectra I(E)/n(E) is quite acceptable. In contrast, the range 700-1200 eV shows larger differences on the high $E_K$ side of the photoelectron peaks, which signals internal electron diffraction [**17**].

The difference between the NPL-IERF and the individual curves I(E)/n(E) in not a problem in the present context, because they have no influence on the final conclusions.

II-2-2  Repoux et alii

Eq.[1] can be readily applied to our experimental conditions by setting the focusing voltage to its appropriate value VF1= 0.8 V which yields $TD_{REPOUX}(E_K) = 202.36 E_K^{1.1956}$.

II-2-3  Bias method

The principle of the so-called Bias-Method (BM in the following) as described by L. Zommer [7] is as follows. The sample is set to a constant bias dc voltage $V_{bias}$ during analysis. A photoelectron emitted from its surface with kinetic energy $E_K$ is detected by the spectrometer with the kinetic energy $E_K^* = E_K + eV_{bias}$. In a plot versus $E_K^*$, the photoelectron spectra recorded at different bias voltages appear shifted along the kinetic energy scale. In a plot versus $E_K = E_K^* - eV_{bias}$ they show a variable difference in intensity only (**Fig.2**). The transmission function and the photoelectron intensity value $I$ are related by :

$$[6] \quad \left.\frac{d\ln I}{dV_{bias}}\right|_{E_K} = \left.\frac{d\ln TD(E_K)}{dE_K}\right|_{E_K}$$

From a set of $I$ and $V_{bias}$ values at a given $E_K$, Zommer evaluated the LHS of Eq.[6] at $V_{bias} = 0$ at different $E_K$ and then fitted a polynomial $p(E_K)$ to the results. The final step consisted in integrating Eq.[6] according to :

$$[7] \quad \frac{TD(E_K)}{TD(E_K^0)} = \exp\left(\int_{E_K^0}^{E_K} p(E_K)dE_K\right)$$

In our investigation we found it realistic and computationally practical to fit a decreasing exponential function to the $I(V_{bias})$ data :

$$[8] \quad I(V_{bias}) = A\exp(-b \times V_{bias})$$

and next to proceed to the numerical integration of the values of parameter $b$ with respect to $E_K$. The variation of $b$ with $E_K$ is shown in **Fig.3**.

One point worth mention here is that the correlation factor of the exponential fit to the graphs $I$-vs-$V_{bias}$ decreases gradually with increasing $E_K$. Above say $E_K$=500eV $R^2$ values lower than 0.8 appear. Above about 750eV, $R^2$ values between 0 and 0.8 can be observed and only the general trend towards a zero limit value (corresponding to a constant transmission function) seems to be physical consistent).

The dependence of $TD_{BM}(E_K)$ on $E_K$ appears in **Fig.4**. Obviously, the best fit to the data points is not a simple power function $E_K^n$. An additional constant term is required to adequately represent the experimental data and the following equation will be used in the subsequent application:

[9]    $TD_{BM}(E_K) = 0.175 + 774 E_K^{1.46}$

In contrast to Eq.[1], Eq.[9] is determined for a single VF1 value equal to 0.8V.

## III. Results

As a representative example, **Fig.5** illustrates the processing of the raw spectrum of gold. One possible evaluation of the true spectrum was first obtained by division of the raw spectrum by the transmission-detection function given by **Eq.[1]** (upper spectrum). Next, the secondary electrons background was computed. In order to reduce the accumulation of errors in the integration process resulting from the signal noise, the computation was carried out over four broad energy windows instead of the entire spectrum. The difference (lower spectrum) represents one possible evaluation of the PES of gold.

The spectra of all three Cu, Ag and Au samples were corrected in a similar way using successively the 3 evaluations of the transmission detection function of Section III (actually, normalized values of the transmission-functions equal to $TD(E_K)/TD(1000)$ were used). Next the RTF values were computed by the FPM method and the results are shown in graphical form in **Fig.6-8** as a function of $E_K$. The transmission detection-function used and the parameters of the linear regression analysis of the data are indicated in each figure. Similar results are presented in **Fig.9** and **Fig.10** that correspond to the assumption of a constant transmission function equal to unity and to the NPL PES obtained as explained in section II respectively. In this latter case, it should be noted that the parameters $Q$, $\beta_{eff}$ and $cos^2\psi$ have not be modified to account for the different $\alpha$ values of the MSPEC but the influence on the results is certainly small.

The coefficient of determination $R^2$ of the linear regressions is a first statistical estimator of the merit of the transmission-detection function used: the lower the former the "better" the latter. Of course, the other steps of the processing of the spectra by the FPM also contribute but at minor level. The observed hierarchy of the $R^2$ values is :

Repoux et alii > IERF NPL > Reference NPL Spectra >> Polarisation > TD(E)=1

In the present context, another pertinent statistical characteristics is the dispersion around the mean: it carries information the accuracy to be expected in practical XPS analysis of mixed compounds based on the combined use of one particular transmission-detection function with the modern version of the FPM method. One simple estimator of dispersion may be the variation coefficient (ratio of the standard deviation around the mean over the mean). In this case the following hierarchy of performances is observed :

Reference NPL Spectra > Repoux et allii ≈ IERF NPL > Bias method > TD(E)=1

In practical quantitative XPS analysis, the composition usually is computed from the area of the most intense emission line of each element. In the present case we may simulate the performance of the different transmission-detection functions in the analysis of a hypothetical homogeneous mixture of the 3 elements Cu, Ag Au as follows. The RTF of these 3 particular lines in the individual spectra should ideally be equal and the relative fraction of each in the sum to 33%. When this figure is compared to the actual results (see Table1) the same hierarchy as above is observed.

## IV Conclusions

It is clear from the foregoing that the MAC2 analyzer cannot reach the performance of the NPL-MSPEC in a quantitative XPS analysis. The estimators of the dispersion of the RTF values are roughly 50-60% lower with the latter compared to the best result obtained with the former. The best performance of the MAC2 has been obtained owing to the combination of the transmission-detection function of Repoux et al. (Eq.[1]) with Tougaard's secondary electron background modelling and use of the modern formalism for XPS analysis taking the theory of transport into account. The superiority over the transmission-detection function of Repoux et alii over that evaluated by the NPL system is narrow if not insignificant. The reverse situation would be a priori expected but the particular shape of the NPL IERF is evidence of a significant contribution of secondary electrons emitted from the inner wall of the cylindrical mirror (in both stages) to the signal intensity delivered by the MAC2 operated in the standard conditions mentioned ($E_{pass}$ = 0.8eV). Indeed, based on detailed explanations by Seah and Smith [**16-17**] of the similar situation occurring in an Auger electron spectrum, one may say that the scattered electrons associated with an intense photoelectron peak should generate a rising step in the in the plot of I(E)/n(E) at kinetic energies above the peak. This is what appears in Fig.1 for the intense XPS peaks of the three metals and also for the strongest Auger peak of Cu.

## Appendix A: Modern formalism for quantitative XPS

We are dealing here with the quantitative analysis of a supposedly perfectly flat surface of a dense single-phase solid irradiated by an unpolarised X-ray beam. Based on the common formalism for XPS (in which elastic scattering is neglected) the signal intensity (spectrometer output) for any given photoelectron line can be expressed as :

$$[A1] \quad I^X = TD(E_K) \int_{z=0}^{\infty} N_V^X \Phi \frac{A_0}{\cos\theta} \left(\frac{d\sigma^X}{d\Omega}\right) \Delta\Omega e^{-\frac{z}{\lambda^X \cos\alpha}} dz = TD(E_K) N_V^X \Phi \frac{A_0}{\cos\theta} \left(\frac{d\sigma^X}{d\Omega}\right) \Delta\Omega \lambda^X \cos\alpha$$

where $TD(E_K)$ represents the spectrometer composite transmission-detection function whose dependence on electron kinetic energy $E_K$ constitutes the main subject of the present paper. In Eq.[A1] the variable z measures the depth below the solid surface, $N_V^X$ is the atomic concentration (number of atoms per unit volume) of element X in the solid, $\Phi$ is the incident flux of photons assumed not attenuated in the subsurface, $A_0$ is the surface area actually seen by the spectrometer collector of acceptance $\Delta\Omega$ (solid angle), $\theta$ is the angle between the normal to the surface and the CMA revolution axis, whereas $\alpha$ is the emission angle (i.e. between the normal to the surface and the direction of the collected photoelectrons.

The term $d\sigma^X/d\Omega$ is the differential photoelectric (or ionization) cross-section. In the present context (unpolarised X-ray beam) it writes :

$$[A2] \quad \frac{d\sigma^X}{d\Omega} = \frac{\sigma_{nl}^X}{4\pi}\left[1 - \frac{\beta}{4}(3\cos^2\psi - 1)\right]$$

where $\sigma_{nl}^X$ is the total photoelectric cross-section of the atomic subshell nl, $\beta$ is the atomic asymmetry parameter and $\psi$ the angle between the incident X-ray beam and the directions of the detected photoelectrons. In the CMA based MAC2 spectrometer, the photoelectrons are collected in a small solid angle $\Delta\Omega$ limited by two coaxial cones. As a consequence $\psi$ varies over a relatively broad range of values and the term $\cos^2\psi$ in Eq.[A2] should be replaced by its average value over $\Delta\Omega$ (see Appendix B).

The modern formalism for XPS takes into account the elastic scattering of the photoelectrons within the solid. In practice this is done in the frame of the so-called transport approximation [18] by modification of Eq.[A2]. An effective asymmetry parameter $\beta_{eff}$ is substituted for $\beta$ and an additional correction parameter, $Q$, is introduced, according to :

$$[A3] \quad \frac{d\sigma^X}{d\Omega} = \frac{\sigma_{nl}^X}{4\pi} Q^X \left[1 - \frac{\beta_{eff}^X}{4}(3\cos^2\psi - 1)\right]$$

It is important to realize that both $Q$ and $\beta_{eff}$ are now dependent on the chemical composition of the solid in which the emitting atom X is embedded through the transport mean free path $\lambda_{tr}$ (also noted TRMFP). The following equations relate these various parameters [19]:

$$[A4] \quad Q = (1-\omega)^{1/2} H(\cos\alpha, \omega)$$

$$[A5] \quad \beta_{eff} = \beta \frac{1-\omega}{Q}$$

$$[A6] \quad \omega = \frac{\lambda}{\lambda + \lambda_{tr}}$$

The combinaison of Eq.[A3,A4,A5] yields :

$$[A7] \quad \frac{d\sigma^X}{d\Omega} = \frac{\sigma_{nl}^X}{4\pi}\left[(1-\omega^X)^{1/2} H(\cos\alpha, \omega) - \frac{\beta^X}{4}(1-\omega^X)(3\cos^2\psi - 1)\right]$$

Here $H$ represents the Chandrasekhar function whose numerical evaluation is somewhat difficult. According to Jablonski et Tougaard [20] the best approximation of $H$ over the range $\omega$ values of interest in XPS writes:

$$[A8] \quad H(\cos\alpha, \omega) = \frac{1 + 1.90781\cos\alpha}{1 + 1.90781\cos\alpha(1-\omega)^{1/2}}$$

In order to proceed through the calculations, on first needs to know the numerical values of $\lambda$ and $\lambda_{tr}$. The IMPF values ($\lambda$) were obtained from the NIST database [21]. For the present application to Cu, Ag and Au the available option "recommended IMPF values" was preferred. The TRMFP was computed using Eq.[A9] below and the numerical values of the elastic scattering cross section $\sigma_{tr}$ available from the NIST database [22] :

$$[A9] \quad \lambda_{tr}^X = \frac{1}{N_V^X \sigma_{tr}^X}$$

These last three parameters are varying with the kinetic energy of the photoelectron, and for each emission line they were evaluated at the measured peak energy. Finally, the numerical values of the atomic asymmetry parameter and of the total photoelectric cross-section were read from the tables of Yeh and Lindau [23] and Scofield [24] respectively.

In conclusion, the $TD(E_K)$ can be expressed in the frame of the first principles method, by the following equation

$$[A10] \quad TD(E_K^X) = \frac{I(E_K^X)}{K \times N_V^X \lambda^X \cos\alpha \left(\frac{d\sigma^X}{d\Omega}\right)}$$

$K = \Phi \frac{A_0}{\cos\theta} \Delta\Omega$ is a constant factor that is usually not evaluated, in which case the energy dependence of the transmission function only can be measured.

## Appendix B

The sketch in **Fig.11** defines the geometrical parameters. **U** is a unit vector along the direction of propagation of an X-ray photon incident in point O, **V** a unit vector along the CMA revolution axis and **W** a unit vector along one propagation direction of the collected photoelectron, i.e. along one generating line of the cone with half angle $\alpha$. The angle between **U** and **V** is $\pi$-$\beta$ with $\beta$ equal to 58° in our experimental setting-up. The cosine of angle $\psi$ between **U** and **W** is equal to the dot product **U**•**W**. The cartesian coordinate system Ox,Oy,Oz is so defined as to have **V** colinear with Oz, and **U** in plane xOz. Using spherical coordinates $r,\theta,\phi$ one may write : $\mathbf{U} = \mathbf{i}\sin\beta - \mathbf{k}\cos\beta$ and $\mathbf{W} = \mathbf{i}\sin\alpha\cos\theta + \mathbf{j}\sin\alpha\sin\theta + \mathbf{k}\cos\alpha$. Hence it can be readily shown that :

$$[B1] \quad \cos\psi = \sin\beta\sin\alpha\cos\theta - \cos\beta\cos\alpha$$

We are interested in the average value of $\cos^2\psi$ over the spectrometer acceptance solid angle:

$$[B2] \quad \overline{\cos^2\psi} = \frac{1}{2\pi(\alpha_{max} - \alpha_{min})} \int_{\theta=0}^{2\pi} \int_{\alpha=\alpha_{min}}^{\alpha_{max}} (\sin\beta.\sin\alpha.\cos\theta - \cos\beta.\cos\alpha)^2 d\alpha\, d\theta$$

The analytic solution of the double integral in Eq.[B2] was evaluated for $\alpha_{min} = 24°$, $\alpha_{max} = 38°$ and $\beta = 58°$ and one finally obtained :

[B3] $\quad \overline{\cos^2 \psi} = 0.3019$

This result was used together with Eq.[A7] to compute the numerical value of the differential ionization cross-section.

The relative error that would result from neglecting the distribution of $\alpha$ values between $\alpha_{min}$ and $\alpha_{max}$ and replacing $\alpha$ by its mean value $\overline{\alpha} = 31°$ amounts to 0.06% only. Thus, in practice and to a very good approximation Eq.[B2] may be reduced to:

[B4] $\quad \overline{\cos^2 \psi} = \dfrac{1}{2\pi} \int\limits_{\theta=0}^{2\pi} (\sin\beta.\sin\overline{\alpha}.\cos\theta + \cos\beta.\cos\overline{\alpha})^2 \, d\theta = \dfrac{1}{2}\sin^2\beta.\sin^2\overline{\alpha} + \cos^2\beta.\cos^2\overline{\alpha}$

Notice that the relative error that would result from replacing $\overline{\cos^2 \psi}$ by $\cos^2\beta$ in Eq.[A7] can also be computed and is found in the range 1.3-2.4%.

**Table 1**:

Statistical parameters of the energy dependence and variability of the RTF values measuring the potential performance of the various evaluations of the transmission-detection function in applied quantitative XPS analysis. Underlined boldface figures signal the best performance for each estimator.

|  | Ideal | $TD_{NPL}(E_K)$ | $TD_{Repoux}(E_K)$ | $TD_{BM}(E_K)$ | $TD(E)=1$ | NPL ref. spectra |
|---|---|---|---|---|---|---|
| Mean | NA | 6561 | 42490 | 26292 | 1464345 | 10578 |
| Standard deviation | 0 | 1624 | 10456 | 10110 | 770050 | 1281 |
| CV (coeff. of variation) | 0 | 0.25 | 0.25 | 0.38 | 0.53 | **<u>0.12</u>** |
| $R^2$ (Coeff of determ.) | 0 | 0.09 | **<u>0.02</u>** | 0.74 | 0.71 | 0.14 |
| | | | | | | |
| Min | NA | 4189 | 27213 | 12532 | 626699 | 8907 |
| Max | NA | 10932 | 64942 | 48757 | 3109530 | 13128 |
| (Max-Min)/Mean | 0 | 1.03 | 0.89 | 1.38 | 1.70 | **<u>0.40</u>** |
| | | | | | | |
| Composition of virtual mix. | | | | | | |
| Cu% | 33 | 45.5 | 38.8 | 52.2 | 58.2 | 32.7 |
| Ag% | 33 | 28.7 | 30.9 | 27.2 | 24.0 | 37.0 |
| Au% | 33 | 25.8 | 30.3 | 20.6 | 17.8 | 30.3 |
| Mean deviation from ideal | 0 | 8.1 | 3.6 | 37.8 | 49.8 | **<u>2.4</u>** |
| largest | 0 | 12.2 | 5.4 | 18.9 | 24.9 | **<u>3.7</u>** |
| lowest | 0 | 4.6 | 2.5 | 6.2 | 9.4 | **<u>0.7</u>** |

**Figure 1** : Graph of the intensity energy response function generated by the NPL calibration system (dashed curve). The result of the division of our raw experimental Cu, Ag and Au spectra by the true spectra of the same elements derived from the NPL reference spectra are shown for comparison.

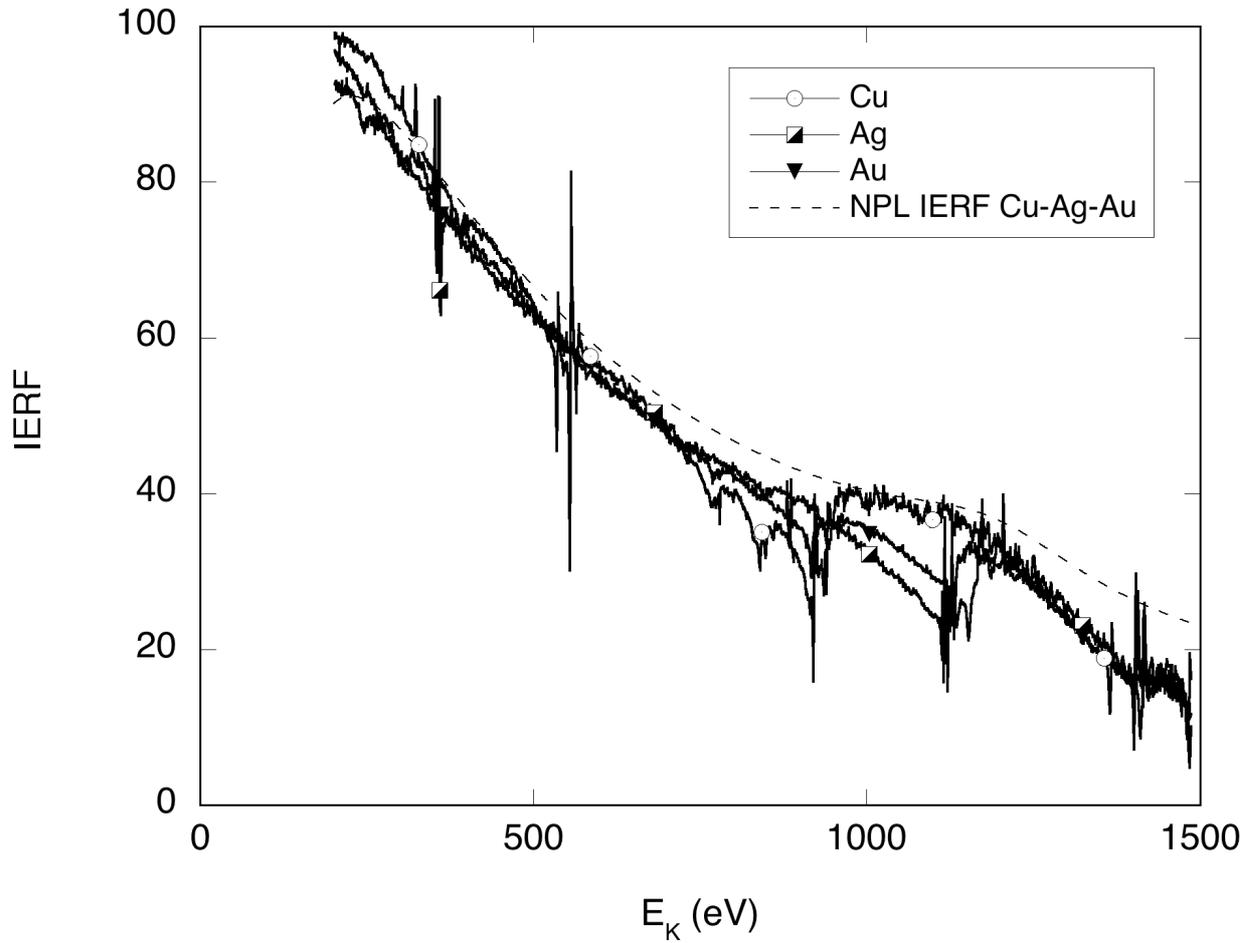

**Figure 2:** Intensity variation of the low $E_K$ section of the raw Ag spectrum recorded with the MAC2 analyzer and various polarisation voltages applied to the sample.

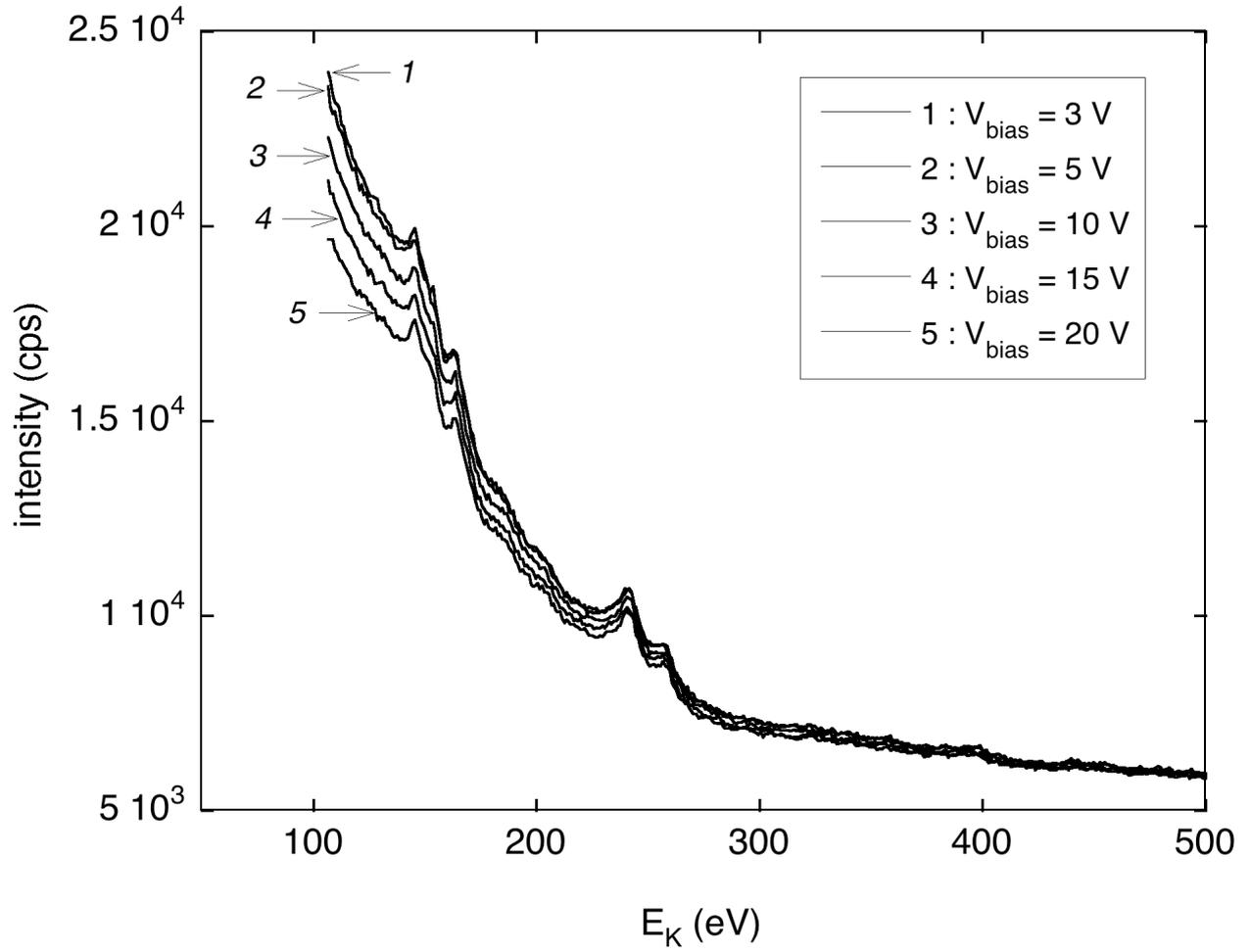

**Figure 3** : Energy dependence of parameter b of Eq.[8]

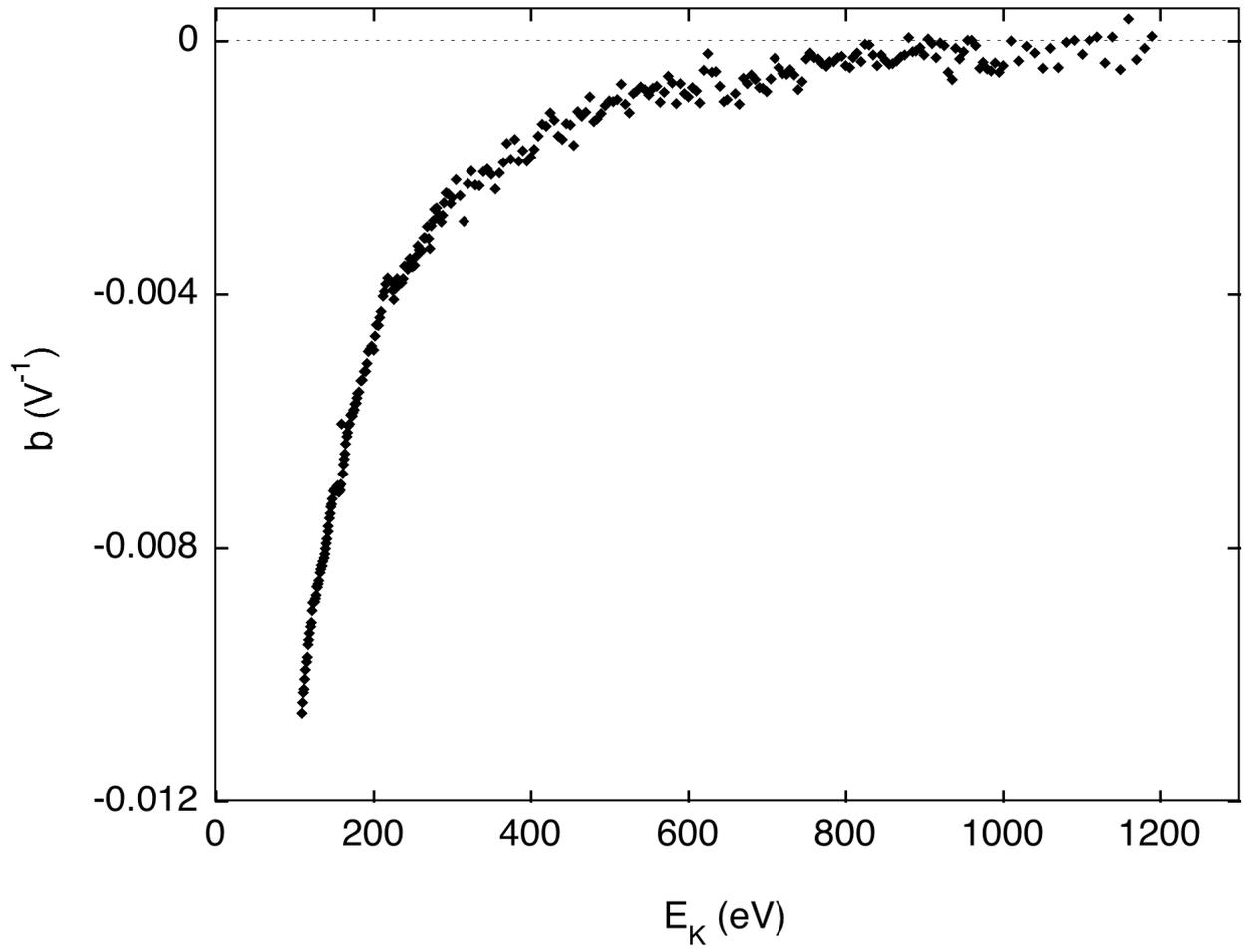

**Figure 4** : Relative transmission-detection function determined by the bias method and best least square fits of power functions to the data.

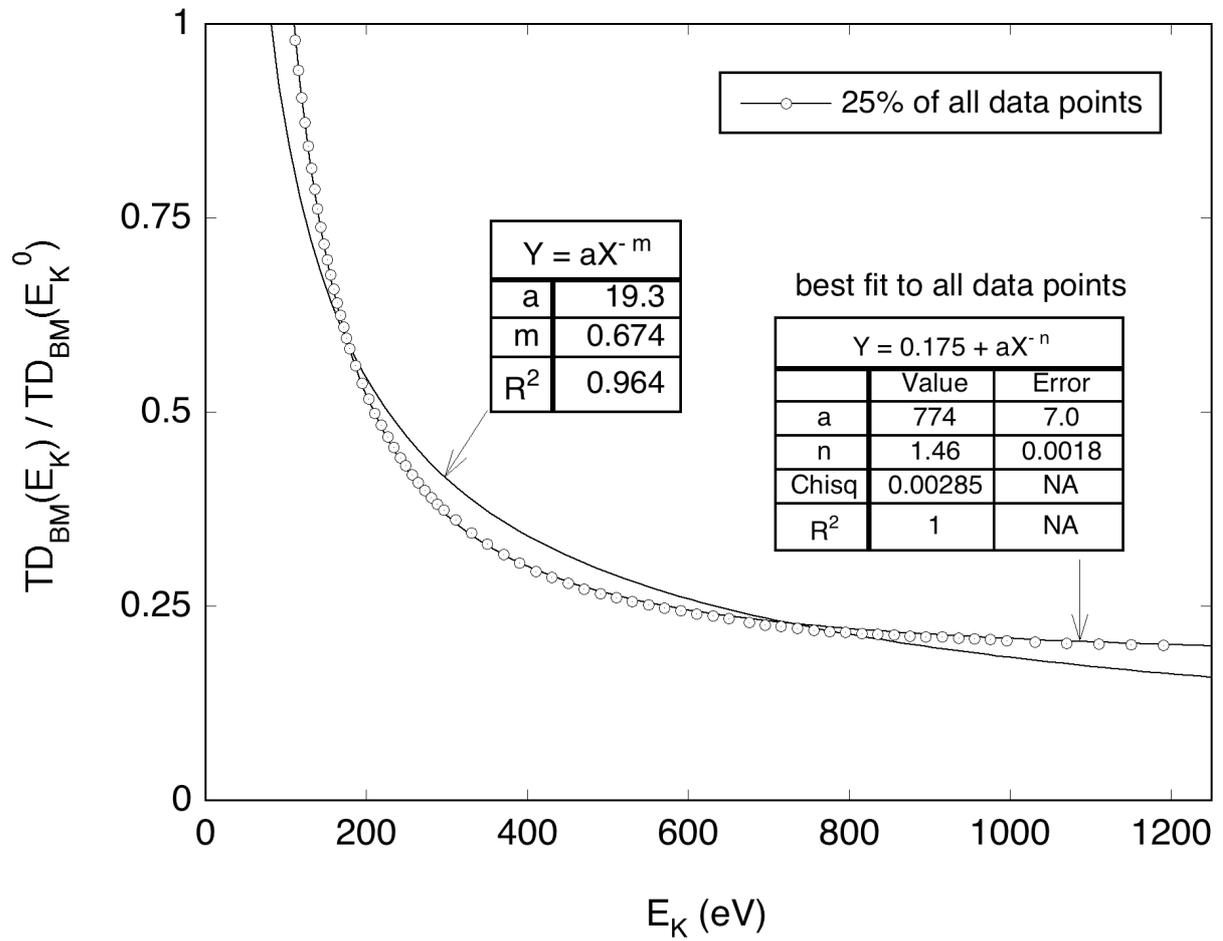

**Figure 5** : Raw photoelectron spectrum of Ag divided by the transmission-detection function $TD_{REPOUX}(E_K)$ (upper spectrum). The primary electron spectrum (lower spectrum) is obtained after subtraction of Tougaard secondary electron background over the indicated energy windows.

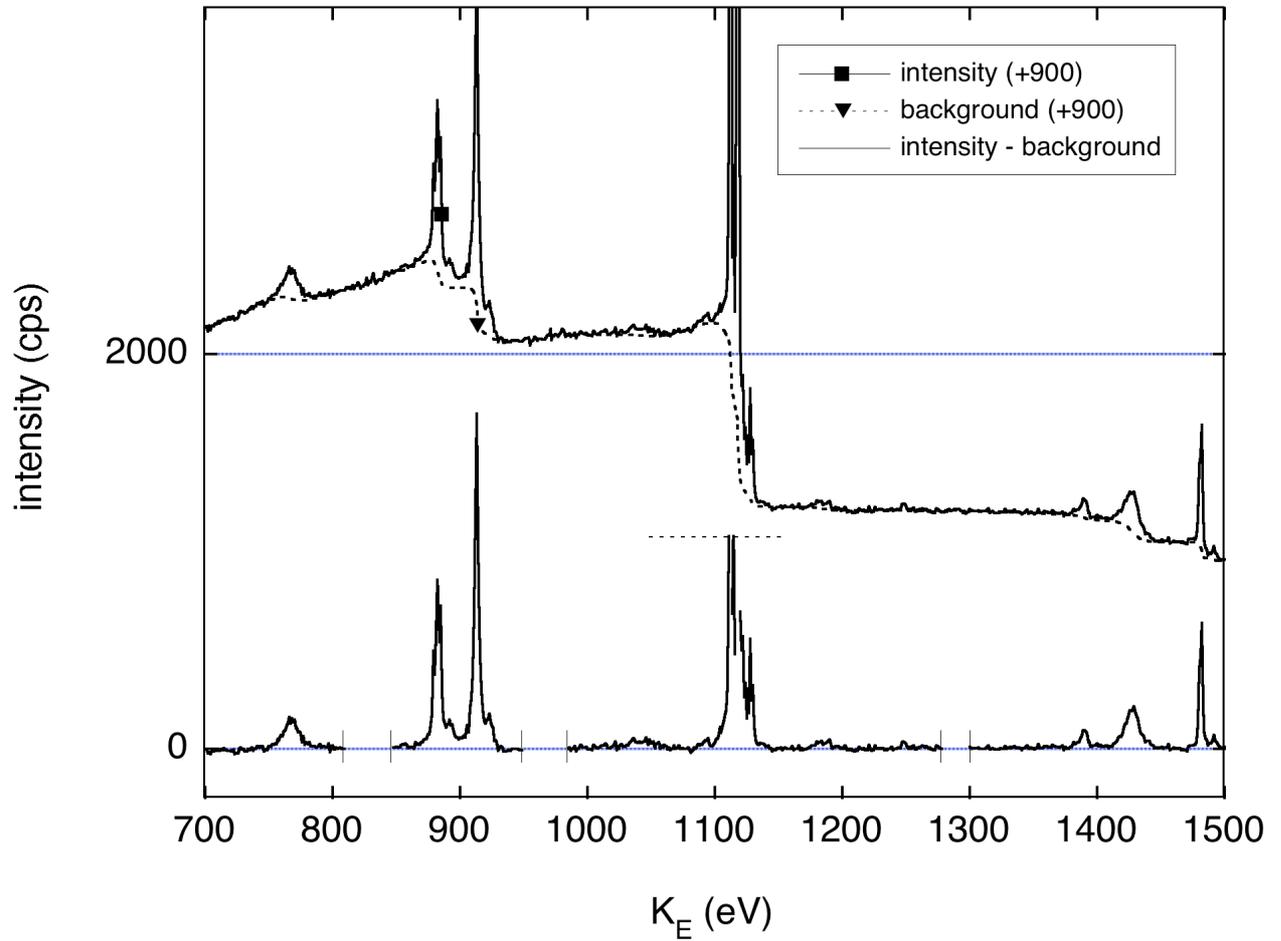

**Figure 6** : RTF vs $E_K$ plot derived from the Cu, Ag and Au spectra initially corrected for the IERF generated by the NPL calibration system

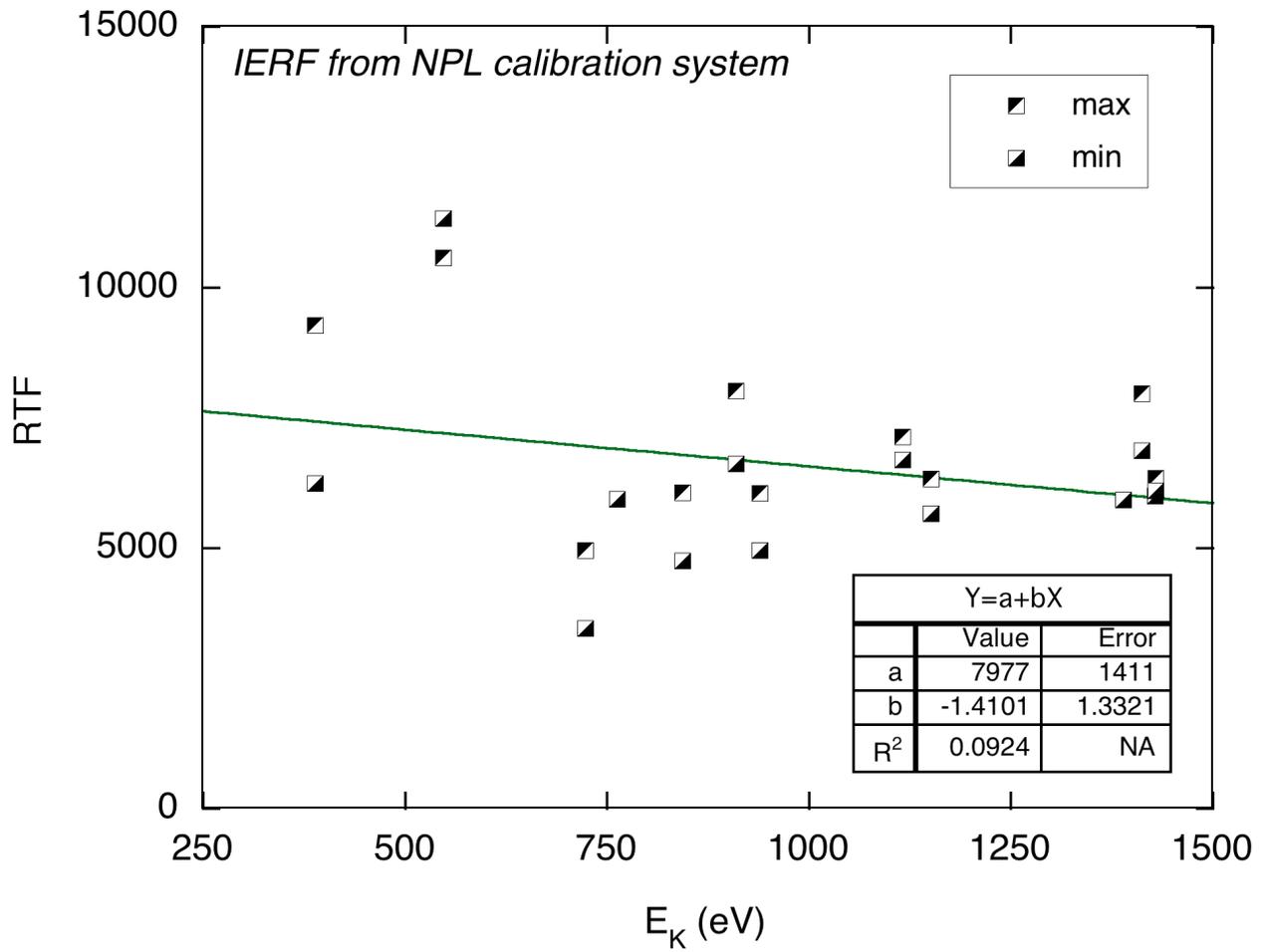

**Figure 7** : RTF vs $E_K$ plot from the Cu, Ag and Au spectra initially corrected for the transmission-detection of Repoux et al.

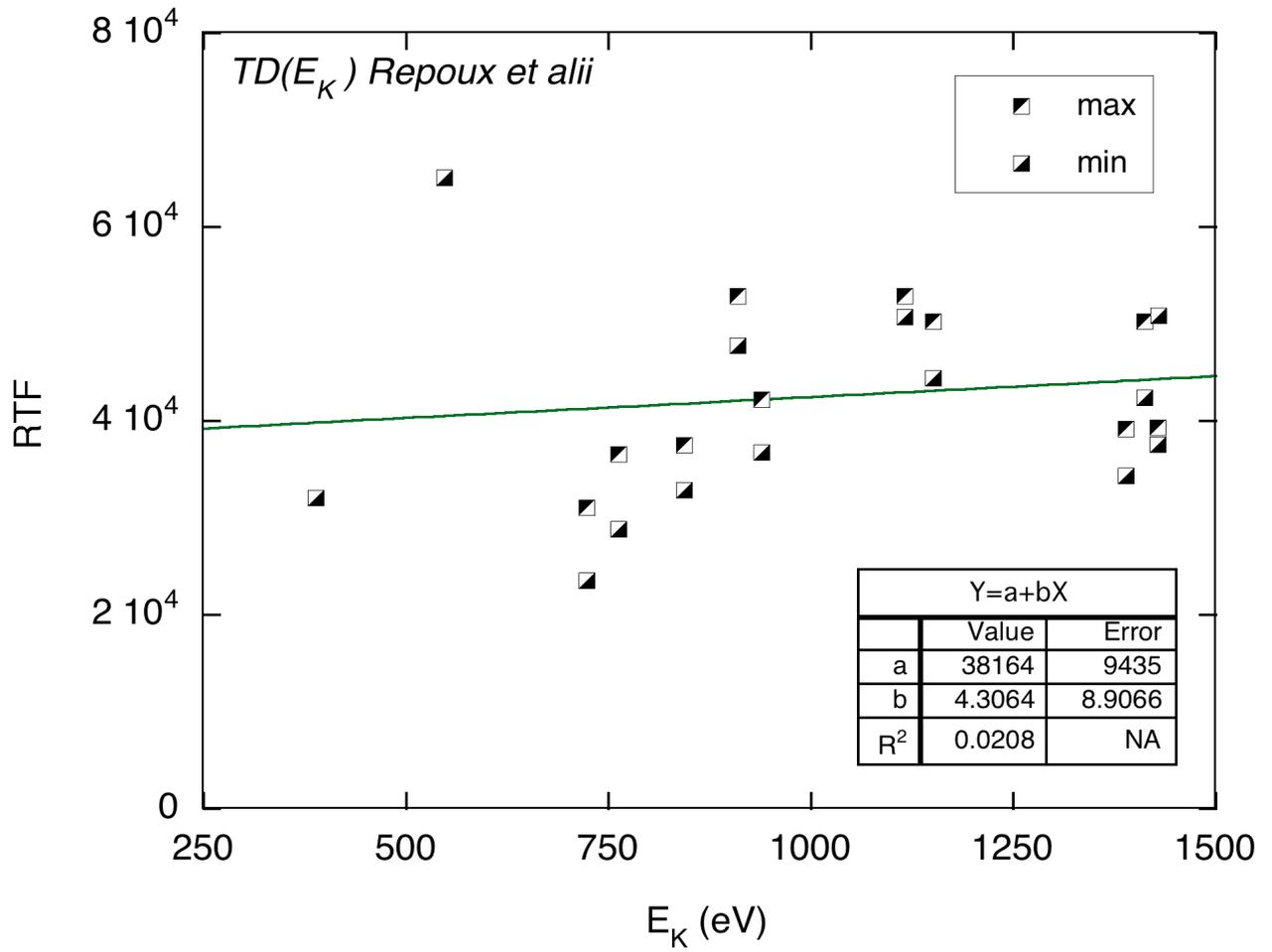

**Figure 8** : RTF vs $E_K$ plot from the Cu, Ag and Au spectra initially corrected for the transmission-detection obtained by the bias method.

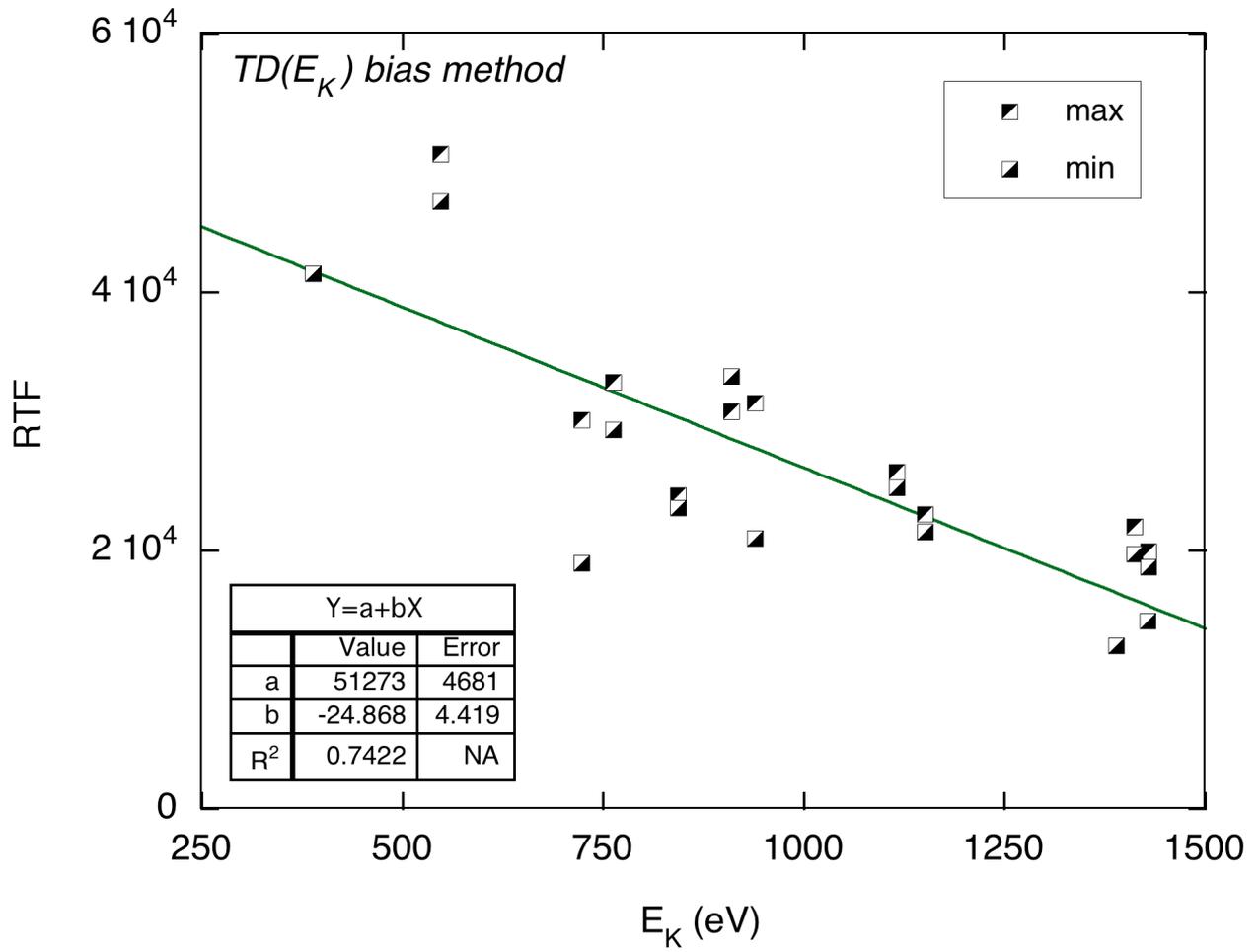

**Figure 9** : RTF vs $E_K$ plot from the raw Cu, Ag and Au spectra.

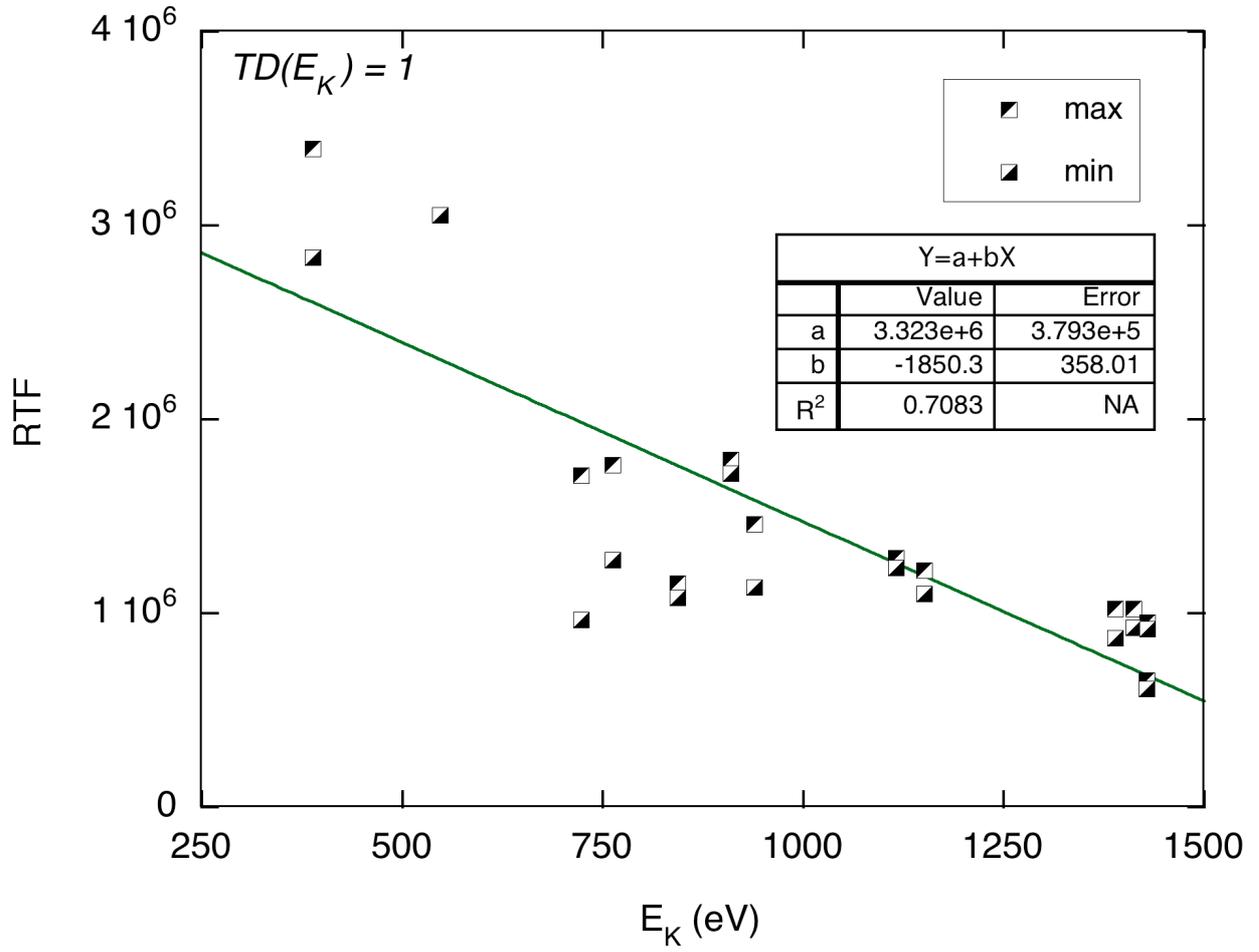

**Figure 10** : RTF vs $E_K$ plot from the Cu, Ag and Au NPL PES.

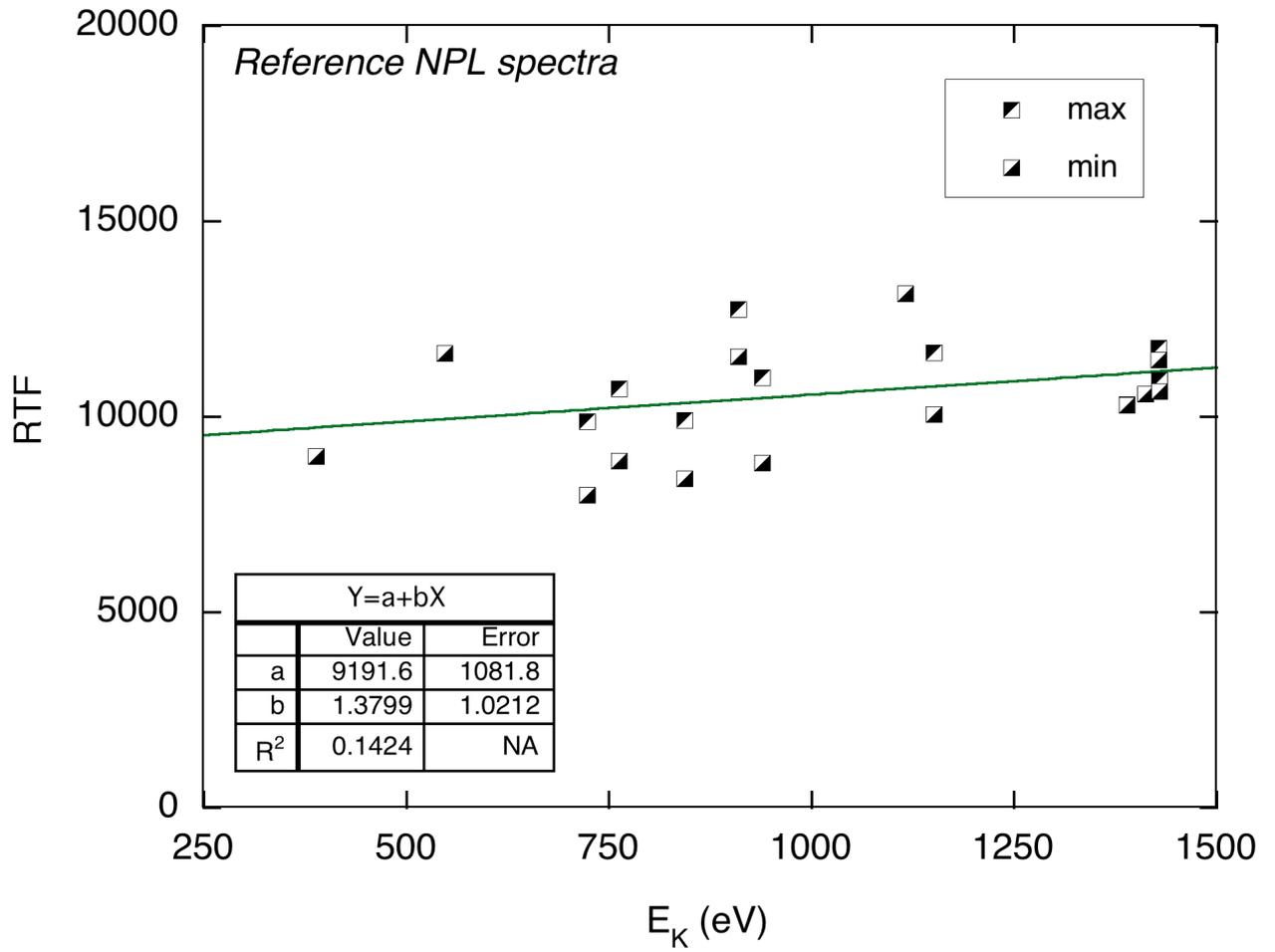

**Figure 11** : schematic diagram showing the angular relations between the direction of the incident X-ray beam, the emitted photoelectron trajectories and the revolution axis of the electron spectrometer perpendicular to the sample surface.

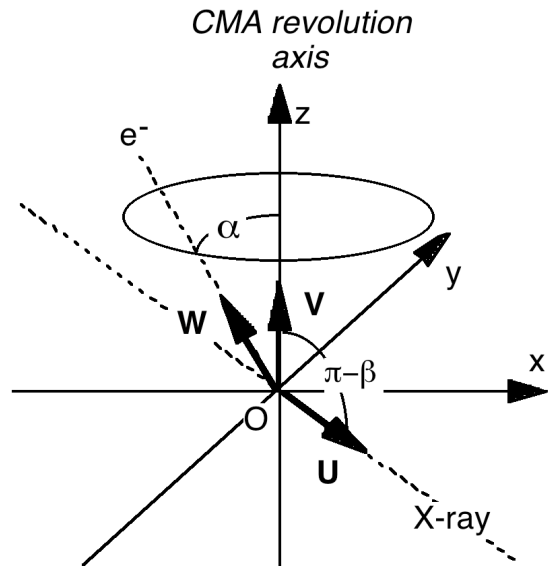